\documentclass[sigconf,nonacm]{acmart}
\AtBeginDocument{%
  }

\setcopyright{none}

\usepackage[]{todonotes}
\usepackage{minted}
\usepackage{wrapfig}
\usepackage{subcaption}
\usepackage{subcaption}
\usepackage{xspace}
\usepackage{enumitem}

\newcommand{\toolname}{\textsc{Maven-Lockfile}}
\newcommand{\etal}{\textit{et al.}\xspace}

\usepackage{regexpatch}
\makeatletter
\xpatchcmd{\@todo}{\setkeys{todonotes}{#1}}{\setkeys{todonotes}{inline,#1}}{}{}
\makeatother

\begin{document}

\title{Maven-Lockfile: High Integrity Rebuild of Past Java Releases}

\author{Larissa Schmid}
\author{Elias Lundell}
\author{Martin Monperrus}
\email{{lgschmid,ellundel,monperrus}@kth.se}
\affiliation{%
  \institution{KTH Royal Institute of Technology}
  \country{Stockholm, Sweden}
}

\author{Yogya Gamage}
\author{Benoit Baudry}
\email{first.lastname@umontreal.ca}
\affiliation{%
  \institution{Université de Montréal}
  \country{Montréal, Canada}
}

\renewcommand{\shortauthors}{Larissa Schmid, Elias Lundell, Yogya Gamage, Benoit Baudry, and Martin Monperrus}

\begin{abstract}
Modern software projects depend on many third-party libraries, complicating reproducible and secure builds. 
Several package managers address this with the generation of a lockfile that freezes dependency versions and can be used to  verify the integrity of dependencies. Yet, Maven, one of the most important package managers in the Java ecosystem, lacks native support for a lockfile. 
We present \toolname{} to generate and update lockfiles, with support for rebuilding projects from past versions. 
Our lockfiles capture all direct and transitive dependencies with their checksums, enabling high integrity builds. 
Our evaluation shows that \toolname{} can reproduce builds from historical commits and is able to detect tampered artifacts. 
With minimal configuration, \toolname{} equips Java projects with modern build integrity and build reproducibility, and fosters future research on software supply chain security in Java. 
\\ Demonstration: \url{https://youtu.be/eGgR3toBgxU}
\end{abstract}



\maketitle

\section{Introduction}

Modern software projects depend on a large number of third-party libraries. These dependencies make development faster, but they also introduce new risks \cite{ladisa2023sok}. 
In the face of these risks, developers need tools to verify the integrity of dependencies, in order  to  have reproducible build results, avoid incompatible changes between dependency versions, and detect possible tampering of artifacts. 
Without such guarantees, tracking down errors due to wrong dependencies is hard and build results cannot be trusted. 

Lockfiles are a practical way to address these problems. 
They record the exact versions and checksums of direct and transitive dependencies, making builds both reproducible and transparent. 
With lockfiles, developers can confidently compare dependency trees across builds and precisely control the effect of version ranges. 
Prior work shows that developers perceive lockfiles as helpful for improving the efficiency and security of dependency management~\cite{gamage2025designspacelockfilespackage}; for enabling reproducible installs~\cite{kabir2022npmpackages}, for preventing unwanted updates to untested or compromised versions~\cite{williams2025researchdirections}, and for supporting the reliable generation of software bills of materials (SBOMs)~\cite{yu2024correctnesssbom}.
Maven, one of the most widely used build tools in the Java ecosystem, does not provide native lockfile support. This gap limits the reproducibility and security of Maven-based builds, and prevents the research community from investigating high-integrity builds for Java.

We here present \toolname{}, which adds state-of-the-art lockfile support for Maven. Our tool generates lockfiles, validates and rebuilds projects based on them, bringing essential integrity features to Maven. Lockfiles produced by \toolname{} contain versions of dependencies and their checksums, covering both direct and transitive dependencies. \toolname{} is compatible with continuous development practices through a GitHub Action that automates the generation, validation, and update of lockfiles in Continuous Integration pipelines.

\noindent
To sum up, our contributions are: 
\begin{itemize}[leftmargin=*]
    \item A blueprint architecture for high-integrity builds in the context of Maven -- one of the most important build systems in enterprise contexts. 
    \item \toolname{}, a tool\footnote{https://github.com/chains-project/maven-lockfile/} for ensuring the integrity of a Maven build and for guarding the build against malicious tampering with the artifacts. 
    \item An experimental evaluation showing how \toolname{} enables rebuilding old versions of a non-trivial, real-world Java project. 
\end{itemize}
\section{Lockfiles and Pinning}

Modern software depends on many external libraries, which are managed and fetched using a package manager. To control the versions of these dependencies, one option is to  explicitly fix the version of a direct dependency in the project's manifest file, a practice sometimes referred to as ``pinning''. However, pinning in manifest files has a clear limitation: it does not control the versions of transitive dependencies. These may still change silently, making builds unpredictable and harder to reproduce~\cite{kabir2022npmpackages}.

\textbf{Lockfiles} extend the idea of pinning by freezing the entire dependency graph, recording the exact versions of both direct and transitive dependencies as they are resolved by the package manager ~\cite{gamage2025designspacelockfilespackage}. Lockfiles may also contain additional details, such as the version of the package manager used and checksums for each dependency package.

Lockfiles make builds reproducible and transparent, no matter when or where they are executed. They provide several important benefits: (i) Deterministic builds: The same dependency graph is installed every time the project is built, avoiding silent upgrades~\cite{williams2025researchdirections}, (ii) Integrity and security: Checksums in lockfiles allow the verification of downloaded artifacts, reducing the risk of tampering~\cite{gamage2025designspacelockfilespackage}, (iii) Transparency and auditing: Lockfiles enable comparing builds, enabling a full audit trail of changes in dependencies.

\textbf{Maven} is one of the most popular package managers in the Java ecosystem. In Maven, dependencies are declared in the project's \texttt{pom.xml} file. Each dependency is identified by three attributes: groupId, artifactId, and version. The version may be fixed, but developers can also choose to declare it as the latest available version; or as a version range. 
When building a project, Maven resolves the dependencies of those packages: it determines which specific version to use for every dependency of a project. This process is fundamentally not deterministic. Even if a project's own \texttt{pom.xml} specifies exact versions, the transitive dependencies may still rely on ranges or unspecified versions. As a result, different builds at different points in time may resolve to different versions, leading to builds that behave differently.
This problem is compounded by the fact that resolved dependencies are largely invisible to developers. They are not aware that the versions of transitive dependencies have changed, which makes debugging and reproducing builds more difficult. In addition, Maven does not provide a built-in mechanism to ensure that downloaded artifacts have not been tampered with during distribution. 
These fundamental limitations highlight the need for proper lockfile support in Maven to guarantee determinism, integrity, and security of Maven builds.

\begin{figure}
    \centering
    \includegraphics[width=\linewidth]{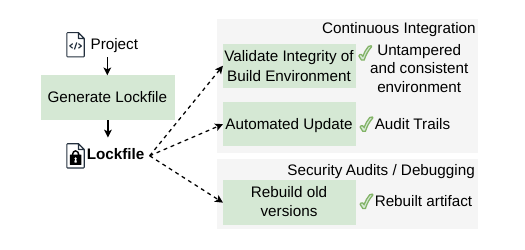}
    \caption{Overview of the unique, novel integrity features of \toolname{}.}
    \label{fig:lifecycle}
\end{figure}

\section{Maven-Lockfile} \label{sec:tool}

\toolname{} brings lockfiles to Maven (cf. Figure~\ref{fig:lifecycle}): It enables generating lockfiles (\autoref{sec:tool:generate-lockfiles}), validating the integrity of build environments based on them (\autoref{sec:tool:validate-integrity}), rebuilding old versions of a project (\autoref{sec:tool:rebuild}), and automatically updating lockfiles (\autoref{sec:tool:automatic-update}), equipping Java projects with high integrity builds (\autoref{sec:tool:novelty}). 

\subsection{Generate Lockfiles} \label{sec:tool:generate-lockfiles}

\toolname{} can generate a lockfile that records the complete dependency graph of a Maven project, including checksums of all artifacts, by executing a single command. This allows developers to freeze and verify dependencies with both direct and transitive coverage. 
The lockfile is written in a human-readable JSON format and contains the full dependency tree, similar to a software bill of materials (SBOM). Figure~\ref{lst:lockfile} shows an excerpt of an example lockfile. \toolname{} supports multi-module projects, creating one lockfile per module. 
For each artifact, the lockfile stores the hashes of all transitive dependencies. This makes it possible to validate not only direct dependencies but also the entire chain of dependencies.

\toolname{} also offers the possibility to add Maven plugins to the lockfile. Maven plugins are themselves artifacts downloaded from external repositories, forming part of the software supply chain and influencing the build result. If a plugin is replaced or tampered with, it could compromise the entire build process. Including plugins in the lockfile ensures that their versions and checksums are validated, protecting not only the project's dependencies but also the integrity of the build system itself. 

The checksums for lockfile generation can be resolved or retrieved using two different modes: 
\textit{Remote} downloads the checksum directly from the Maven repository.
\textit{Local} calculates the checksum directly from the artifact stored in the local Maven cache. 
Remote checksums provide stronger guarantees that the downloaded artifact  is the same as the one published by the repository, but this trust relies on the repository itself not being compromised. Local checksums ensure consistency with the artifact already downloaded, but they cannot detect tampering that happened before the artifact was downloaded.

Moreover, \toolname{} can include environment metadata in the lockfile, which allows warnings when the build environment changes. As a change in the build environment can affect the outcome of a build, recording this information helps explain differences in build results and supports reproducibility across systems.

\begin{figure}
      \scriptsize
\begin{minted}{json}
"dependencies": [  { "groupId": "A", "artifactId": "A", 
                     "version": "1.2", "checksum": "85bc8591c87" }, 
                   { "groupId": "D", "artifactId": "D", 
                     "version": "3.4", "checksum": "f92d72b8720" }, ], 
"mavenPlugins": [],
"metaData": { "environment": { 
  "osName": "Linux", "mavenVersion": "3.8.2", "javaVersion": "21.0.5" }
}, 
"config": { "checksumMode": "local" }
\end{minted}
      \caption{Excerpt of an example lockfile.}
      \label{lst:lockfile}
\end{figure}

\subsection{Validate Integrity of Build Environments} \label{sec:tool:validate-integrity}

After generating a lockfile, \toolname{} can validate the integrity of a Maven repository against an existing lockfile. This ensures that all dependencies used in the build are the same as when the lockfile was originally created, and that the local repository has not been tampered with. 
The validation process checks both versions and checksums of dependencies, covering the entire dependency graph. 
If the \texttt{validate} command runs successfully, it means that the build environment is valid; i.e., all defined dependencies match the lockfile. If the validation fails, it means that one or more dependencies have changed since the lockfile was generated, which could indicate either a silent update or a tampering.
\toolname{} also enables skipping specific modules to disable validation in selected parts of a multi-module project.

By validating dependencies against a lockfile, developers have guarantees that the build environment is untampered and consistent with the reference dependencies, no matter whether it is today or far in the future.

\subsection{Rebuild Old Versions} \label{sec:tool:rebuild}

\toolname{} supports rebuilding projects from a lockfile with the \texttt{freeze} feature, making it possible to reproduce older versions of a build with the same dependency resolution. This is done by generating a new POM file, \texttt{pom.lockfile.xml}, that incorporates all dependency information from the lockfile. 
In the generated POM file, every version of the direct dependencies from the original POM is replaced with the versions recorded in the lockfile. In addition, all transitive dependencies are added to the \texttt{dependencyManagement} section, with both their version and scope taken from the lockfile. In this way, the new POM represents a complete and reproducible dependency specification, complying with Maven dependency resolution semantics.
Once \texttt{pom.lockfile.xml} is created, Maven is invoked with it using the \texttt{-f} flag to rebuild the project exactly as described by the lockfile, including both direct and transitive dependencies.

With this functionality, developers can reliably reproduce historical builds, making it easier to investigate bugs, validate security concerns, or recreate past releases.

\subsection{Automated Update of Lockfile} \label{sec:tool:automatic-update}

To support the integration of lockfiles into development workflows, \toolname{} provides a CI pipeline for validating and updating lockfiles automatically.
The pipeline is implemented as  a Github Action, and works as follows: if a \texttt{pom.xml} or \texttt{lockfile.json} file has been modified in a pull request, the action automatically generates an updated lockfile and adds it as a commit to the pull request, creating a full audit trail of dependency changes. If no changes are detected in these files, the action validates the existing lockfile and fails the build if the lockfile is incorrect (cf. Section~\ref{sec:tool:validate-integrity}). 
Per the best practices for lockfiles, the \toolname{ }action enforces all CI to remain consistent with their declared lockfiles and supports keeping them up-to-date. 

\subsection{Novelty of \toolname{}} \label{sec:tool:novelty}

When generating a lockfile for a project, one needs to include four key fields \cite{gamage2025designspacelockfilespackage}: resolved package versions, package checksums, package source, and a way to distinguish between resolved direct and indirect dependencies . 
Among the most popular package managers, only Go and Cargo include all these elements. 
\toolname{} is the first system to equip Java projects with state-of-the art build integrity. 

The other popular build automation tool for Java, Gradle, includes a built-in solution to generate lockfiles. 
However, in practice, Gradle lockfiles are rarely used because their configuration is not user-friendly. First, the lockfile generation is not the default behavior. 
In addition to limited usability, Gradle lockfiles  omit important details that should be part of a lockfile, such as checksums of resolved dependencies. 
The key novelty of \toolname{} compared to Gradle is that it requires minimal configuration effort from developers and includes the most important elements to ensure integrity and reproducibility.
\toolname{} finally brings high integrity builds to Java developers.
\section{Evaluation} 

We evaluate \toolname{} based on two research questions:

\begin{enumerate}[label=\textbf{RQ\arabic*},leftmargin=*]
    \item Can \toolname{} rebuild old versions thanks to the lockfile?
    \item Can \toolname{} detect tampered dependency artifacts? 
\end{enumerate}

\subsection{Methodology}

To answer RQ1, we use \toolname{} to rebuild ten randomly chosen previous releases of \toolname{} itself, starting from the first release supporting rebuilding from 2023-06-05. 
For each release, we checkout the corresponding commit and the corresponding lockfile. We first reproduce the build environment by downloading the Java and Maven versions specified in the lockfile. 
We then use the chosen release of \toolname{} to produce a frozen POM from the committed lockfile of the old release (cf. Section~\ref{sec:tool:rebuild}). Finally, we validate the build environment and rebuild \toolname{} using the frozen POM.

To answer RQ2, we first generate a lockfile for a project and validate the resulting build based on it (cf. Section~\ref{sec:tool:validate-integrity}). For this, we use the latest version of \toolname{}, version 5.7.1.
Then, we modify one of the dependencies of the project by changing the locally downloaded artifact used for the build through a random binary perturbation. We then validate the environment again based on the lockfile, with the expectation that \toolname{} raises an error.

\subsection{Results}

\definecolor{yescolor}{HTML}{221E1F} 
\definecolor{nocolor}{HTML}{949698} 
\newcommand{\y}{{\textcolor{yescolor}{\checkmark}}}
\newcommand{\n}{{\textcolor{nocolor}{$\times$}}}

\begin{table}
    \centering
    \begin{tabular}{lcccccr}
        \toprule
        Version & Date & Generate & Validate & Rebuild \\ 
        \midrule
        2.1.0 & 2023-06-05 & \y & \y & \y \\ 
        2.2.0 & 2023-06-07 & \y & \y & \y \\ 
        3.0.1 & 2023-06-12 & \y & \y & \y \\
        3.1.0 & 2023-06-12 & \y & \y & \y \\
        4.1.0 & 2023-08-29 & \y & \y & \n \\
        5.1.0 & 2024-04-30 & \y & \y & \y \\
        5.3.2 & 2024-12-19 & \y & \y & \y \\
        5.3.4 & 2024-12-20 & \y & \y & \y \\
        5.5.0 & 2025-04-23 & \y & \y & \y \\
        5.6.2 & 2025-09-10 & \y & \y & \y \\
        \bottomrule
    \end{tabular}
    \caption{Results of generating new lockfiles, validating old lockfiles and rebuilding from them.}
    \label{tab:rebuild-from-lockfile}
\end{table}

Table \ref{tab:rebuild-from-lockfile} shows the results of RQ1. 
All releases contain a valid lockfile, demonstrating that the generation of lockfiles behaves correctly and all could be used to validate the build environment. 
In nine out of ten cases, it is possible to rebuild the project based on this environment. 
Major version 4 fails to rebuild due to a bug in the dependency resolution used by \toolname{}. 
Specifically, \toolname{} incorrectly added transitive dependencies with the \texttt{test} scope to the frozen \texttt{pom.xml}, instead of the same transitive dependency without the \texttt{test} scope. 
Dependencies with the \texttt{test} scope are not available in non-test classpaths and thus, the compilation fails as Maven cannot access the expected classes. 
This was fixed in version 5.1.0, after which rebuilding has been stable.

Validating and rebuilding version 3.0.1 and 3.1.0 required the use of \toolname{} version 2. 
This is because the major version 3.x of \toolname{} added new fields to the lockfile which were not backwards compatible. 
As no dependencies changed with the release of version 3, the lockfile in the repository was still generated by version 2 of \toolname{}, and thus required the older version to operate on them.
Using the latest version 2 release of \toolname{}, validating and rebuilding behaves as expected.

When validating versions up to and including 5.1.0, the lockfiles contained the previous \texttt{-SNAPSHOT} version instead of the version specified in the pom at the release tag. 
Thus, the commit before the release tag was used when validating.
Version 5.5.0, similar to major version 3, introduced new fields requiring the use of the previous version of \toolname{} (in this case 5.4.2) to successfully validate.
Moreover, prior to version 4.1.0, optional parameters such as inclusion of maven plugins have to be specified manually for validation.

For RQ2, generating the lockfile for \toolname{} itself and running \texttt{validate} against it gives no errors, which is a prequisite. 
We then modify the first dependency listed in the lockfile, the \texttt{jar} for \emph{com.google.code.gson:gson:2.13.2} by extracting and repackaging it, thus changing access times in the zip. 
Running \texttt{validate} against the lockfile again, \toolname{} correctly raises an error, reporting that the checksum of the \texttt{gson} dependency has changed with respect to the lockfile. 
After rebuilding using the modified dependency, the test suite on the newly built artifact passes, illustrating that some adversarial malicious modification of \texttt{gson} can go undetected under normal Maven operations of a CI/CD pipeline. 

Overall, our evaluation shows that \toolname{} enables rebuilding old releases with reproduced dependency resolution, as frozen in the lockfile.

\subsection{Discussion}

There is scarce literature on longitudinal build reproducibility. We know some factors that break reproducibility, such as missing version information or nondeterministic transitive dependencies, but we lack tools for verifying reproducibility.
To the best of our knowledge, \toolname{} is the first system to enable systematic longitudinal reproducibility in real-world Java projects.

Moreover, we have shown that \toolname{} detects artifacts that have been tampered with. 
Future research can explore how lockfiles can be extended and integrated with existing security mechanisms, such as digital signatures of artifacts or transparency logs, to provide even stronger guarantees for the software supply chain.

Beyond Java, the design of \toolname{} provides a blueprint for bringing lockfile support to Maven and other ecosystems that currently lack it. Our paper informs general strategies for reproducible builds and supply chain protection across different build and dependency management systems.

\section{Related Work}

Several studies mention lockfiles in the context of dependency management and recognize their impact. 
He \etal \cite{he2025pinningisfutile} recommend lockfiles instead of pinning. 
Bogart \etal \cite{Bogart2021} identify the use of lockfiles as a way to stabilize projects by locking indirect dependency versions. 
Venturini \etal \cite{venturini2023dependubreaknpm} show that lockfiles help avoid in-range breaking updates. 
Vaidya \etal \cite{vaidya2019security} suggest that the npm CLI lockfile \texttt{npm-shrinkwrap} can mitigate attacks in the npm ecosystem. 
Kabir \etal \cite{kabir2022npmpackages} consider committing lockfiles one of the best practices. 
To the best of our knowledge, none of these studies do action research as we do: injecting lockfiles support in real-world projects. 
\toolname{} provides a foundation for future research on integrity and build determinism in the context of Java projects.

\section{Conclusion}

We have introduced \toolname{}, a state-of-the-art tool to validate the integrity of Maven builds, reproduce old builds, and automatically integrate with modern software engineering processes. 
Our evaluation shows that \toolname{} can successfully rebuild previous project versions using pinned dependencies from the lockfile, and is able to detect tampered artifacts. 
The novelty of \toolname{} lies in its full coverage of the essential elements needed for integrity and reproducibility, equipping Java projects with modern build integrity at minimal cost for developers. 
Future research is needed to study how lockfiles can integrate with other software supply chain security mechanisms, such as transparency logs.

\begin{acks}
We would like to thank Martin Wittlinger for his contributions to \toolname. 
This work is supported by the CHAINS project funded by the Swedish Foundation for Strategic Research (SSF), and by IVADO and the Canada First Research Excellence Fund. 
\end{acks}

\bibliographystyle{ACM-Reference-Format}
\bibliography{bibliography}


\begin{thebibliography}{9}


\ifx \showCODEN    \undefined \def \showCODEN     #1{\unskip}     \fi
\ifx \showISBNx    \undefined \def \showISBNx     #1{\unskip}     \fi
\ifx \showISBNxiii \undefined \def \showISBNxiii  #1{\unskip}     \fi
\ifx \showISSN     \undefined \def \showISSN      #1{\unskip}     \fi
\ifx \showLCCN     \undefined \def \showLCCN      #1{\unskip}     \fi
\ifx \shownote     \undefined \def \shownote      #1{#1}          \fi
\ifx \showarticletitle \undefined \def \showarticletitle #1{#1}   \fi
\ifx \showURL      \undefined \def \showURL       {\relax}        \fi
\providecommand\bibfield[2]{#2}
\providecommand\bibinfo[2]{#2}
\providecommand\natexlab[1]{#1}
\providecommand\showeprint[2][]{arXiv:#2}

\bibitem[Bogart et~al\mbox{.}(2021)]%
        {Bogart2021}
\bibfield{author}{\bibinfo{person}{Chris Bogart}, \bibinfo{person}{Christian K\"{a}stner}, \bibinfo{person}{James Herbsleb}, {and} \bibinfo{person}{Ferdian Thung}.} \bibinfo{year}{2021}\natexlab{}.
\newblock \showarticletitle{{When and How to Make Breaking Changes: Policies and Practices in 18 Open Source Software Ecosystems}}.
\newblock \bibinfo{journal}{\emph{ACM TSE}} (\bibinfo{date}{July} \bibinfo{year}{2021}).
\newblock


\bibitem[Gamage et~al\mbox{.}(2025)]%
        {gamage2025designspacelockfilespackage}
\bibfield{author}{\bibinfo{person}{Yogya Gamage}, \bibinfo{person}{Deepika Tiwari}, \bibinfo{person}{Martin Monperrus}, {and} \bibinfo{person}{Benoit Baudry}.} \bibinfo{year}{2025}\natexlab{}.
\newblock \bibinfo{title}{The Design Space of Lockfiles Across Package Managers}.
\newblock
\showeprint[arxiv]{2505.04834}~[cs.SE]


\bibitem[He et~al\mbox{.}(2025)]%
        {he2025pinningisfutile}
\bibfield{author}{\bibinfo{person}{Hao He}, \bibinfo{person}{Bogdan Vasilescu}, {and} \bibinfo{person}{Christian K\"{a}stner}.} \bibinfo{year}{2025}\natexlab{}.
\newblock \showarticletitle{Pinning Is Futile: You Need More Than Local Dependency Versioning to Defend against Supply Chain Attacks}.
\newblock \bibinfo{journal}{\emph{Proc. ACM Softw. Eng.}} \bibinfo{volume}{2}, \bibinfo{number}{FSE} (\bibinfo{date}{June} \bibinfo{year}{2025}).
\newblock
\href{https://doi.org/10.1145/3715728}{doi:\nolinkurl{10.1145/3715728}}


\bibitem[Kabir et~al\mbox{.}(2022)]%
        {kabir2022npmpackages}
\bibfield{author}{\bibinfo{person}{Md~Mahir~Asef Kabir}, \bibinfo{person}{Ying Wang}, \bibinfo{person}{Danfeng Yao}, {and} \bibinfo{person}{Na Meng}.} \bibinfo{year}{2022}\natexlab{}.
\newblock \showarticletitle{How Do Developers Follow Security-Relevant Best Practices When Using NPM Packages?}. In \bibinfo{booktitle}{\emph{2022 IEEE SecDev}}. \bibinfo{pages}{77--83}.
\newblock
\href{https://doi.org/10.1109/SecDev53368.2022.00027}{doi:\nolinkurl{10.1109/SecDev53368.2022.00027}}


\bibitem[Ladisa et~al\mbox{.}(2023)]%
        {ladisa2023sok}
\bibfield{author}{\bibinfo{person}{Piergiorgio Ladisa}, \bibinfo{person}{Henrik Plate}, \bibinfo{person}{Matias Martinez}, {and} \bibinfo{person}{Olivier Barais}.} \bibinfo{year}{2023}\natexlab{}.
\newblock \showarticletitle{Sok: Taxonomy of attacks on open-source software supply chains}. In \bibinfo{booktitle}{\emph{2023 IEEE Symposium on Security and Privacy (SP)}}. IEEE, \bibinfo{pages}{1509--1526}.
\newblock


\bibitem[Vaidya et~al\mbox{.}(2019)]%
        {vaidya2019security}
\bibfield{author}{\bibinfo{person}{Ruturaj~K Vaidya}, \bibinfo{person}{Lorenzo De~Carli}, \bibinfo{person}{Drew Davidson}, {and} \bibinfo{person}{Vaibhav Rastogi}.} \bibinfo{year}{2019}\natexlab{}.
\newblock \showarticletitle{Security issues in language-based software ecosystems}.
\newblock  (\bibinfo{year}{2019}).
\newblock
\showeprint[arxiv]{1903.02613}~[cs.CR]


\bibitem[Venturini et~al\mbox{.}(2023)]%
        {venturini2023dependubreaknpm}
\bibfield{author}{\bibinfo{person}{Daniel Venturini}, \bibinfo{person}{Filipe~Roseiro Cogo}, \bibinfo{person}{Ivanilton Polato}, \bibinfo{person}{Marco~A. Gerosa}, {and} \bibinfo{person}{Igor~Scaliante Wiese}.} \bibinfo{year}{2023}\natexlab{}.
\newblock \showarticletitle{I Depended on You and You Broke Me: An Empirical Study of Manifesting Breaking Changes in Client Packages}.
\newblock \bibinfo{journal}{\emph{ACM TSE}} (\bibinfo{date}{May} \bibinfo{year}{2023}).
\newblock


\bibitem[Williams et~al\mbox{.}(2025)]%
        {williams2025researchdirections}
\bibfield{author}{\bibinfo{person}{Laurie Williams}, \bibinfo{person}{Giacomo Benedetti}, \bibinfo{person}{Sivana Hamer}, \bibinfo{person}{Ranindya Paramitha}, \bibinfo{person}{Imranur Rahman}, \bibinfo{person}{Mahzabin Tamanna}, \bibinfo{person}{Greg Tystahl}, \bibinfo{person}{Nusrat Zahan}, \bibinfo{person}{Patrick Morrison}, \bibinfo{person}{Yasemin Acar}, \bibinfo{person}{Michel Cukier}, \bibinfo{person}{Christian K\"{a}stner}, \bibinfo{person}{Alexandros Kapravelos}, \bibinfo{person}{Dominik Wermke}, {and} \bibinfo{person}{William Enck}.} \bibinfo{year}{2025}\natexlab{}.
\newblock \showarticletitle{Research Directions in Software Supply Chain Security}.
\newblock \bibinfo{journal}{\emph{ACM TSE}} \bibinfo{volume}{34}, \bibinfo{number}{5} (\bibinfo{date}{May} \bibinfo{year}{2025}).
\newblock
\showISSN{1049-331X}
\href{https://doi.org/10.1145/3714464}{doi:\nolinkurl{10.1145/3714464}}


\bibitem[Yu et~al\mbox{.}(2024)]%
        {yu2024correctnesssbom}
\bibfield{author}{\bibinfo{person}{Sheng Yu}, \bibinfo{person}{Wei Song}, \bibinfo{person}{Xunchao Hu}, {and} \bibinfo{person}{Heng Yin}.} \bibinfo{year}{2024}\natexlab{}.
\newblock \showarticletitle{On the Correctness of Metadata-Based SBOM Generation: A Differential Analysis Approach}. In \bibinfo{booktitle}{\emph{2024 IEEE/IFIP DSN}}. \bibinfo{pages}{29--36}.
\newblock
\href{https://doi.org/10.1109/DSN58291.2024.00018}{doi:\nolinkurl{10.1109/DSN58291.2024.00018}}


\end{thebibliography}

\end{document}